\documentstyle[eqsecnum,aps]{revtex}

\begin{document}
\draft \preprint{HEP/123-qed}
\title{Path Integral Description of a Semiclassical Su-Schrieffer-Heeger Model}
\author{Marco Zoli}
\address{Istituto Nazionale Fisica della Materia - Universit\'a di Camerino, \\
62032 Camerino, Italy. e-mail: marco.zoli@unicam.it }

\date{\today}
\maketitle
\begin{abstract}
The electron motion along a chain is described by a continuum
version of the Su-Schrieffer-Heeger Hamiltonian in which phonon
fields and electronic coordinates are mapped onto the time scale.
The path integral formalism allows us to derive the non local
source action for the particle interacting with the oscillators
bath. The method can be applied for any value of the {\it e-ph}
coupling. The path integral dependence on the model parameters has
been analysed by computing the partition function and some
thermodynamical properties from $T=\,1K$ up to room temperature. A
peculiar upturn in the low temperature {\it heat capacity over
temperature} ratio (pointing to a glassy like behavior) has been
ascribed to the time dependent electronic hopping along the chain.
\end{abstract}
 \narrowtext
\section*{I. Introduction}

Conjugated organic polymers have been a focus of intensive
research over the last twentyfive years after the discovery of
their remarkable conducting properties tuned by the dopant
concentration \cite{macdiarmid}. Undoped {\it trans}-polyacetylene
sustains neutral solitons in the dimerized chain as a consequence
of the Peierls instability \cite{ssh}. Charge injection in
polymers induces a lattice distortion with the associated
formation of localized excitations, polarons and/or charged
solitons. While the latter can exist only in {\it
trans}-polyacetylene whose ground state is twofold degenerate, the
former solutions are more general since they do not require such
degeneracy. The Su-Schrieffer-Heeger (SSH) model Hamiltonian
\cite{ssh1} is the fundamental tool in the analysis of quasi one
dimensional polymers. In the weak coupling regime a continuum
version \cite{tlm} of the SSH model has been developed and
polaronic solutions have been obtained analytically
\cite{campbell}. Although the 1D properties of the SSH model have
been mainly investigated so far
\cite{schulz,fradkin,stafstrom,raedt,zheng,io}, extensions to two
dimensions were considered in the late eighties in connection with
the Fermi surface nesting effect on quasi 2D high $T_c$
superconductors \cite{tang}. Recent numerical analysis \cite{ono}
have revealed the rich physics of 2D-SSH polarons whose mass seems
to be larger than in the 1D case, at least in the intermediate
regime of the adiabatic parameter \cite{zoli}.

One of the main issues in the study of {\it e-ph} models regards
the possibility to monitor the conditions for polaron formation as
a function of the coupling strength. The Feynman path integral
method \cite{feynman} is a landmark in this respect and several
studies \cite{devreese,farias,ganbold} have applied the path
integral formalism both to the acoustical polaron of the
Fr\"ohlich Hamiltonian  and to the Holstein model \cite{raedt1}.
As a main property of the SSH model, the {\it e-ph} interaction
modifies the bare electron hopping integral of the tight binding
approximation thus leading to an Hamiltonian linearly dependent on
the phonon displacement field. This feature will allow us to
attack the SSH model by path integrals techniques after
introducing a generalized version of the semiclassical model which
treats only the electrons quantum mechanically. The general
formalism is described in Section II together with the method to
compute the full partition function of the interacting system. In
Section III, we focus on the peculiarities of the free energy and
heat capacity as a function of the model parameters. Section IV
contains the final remarks.

\section*{II. Path Integral Formalism}

The 1D SSH interacting Hamiltonian reads:

\begin{eqnarray}
H=\,& & \sum_{r}J_{r,r+1} \bigl(f^{\dag}_r f_{r+1} +
f^{\dag}_{r+1} f_{r} \bigr) \,
 \nonumber \\
J_{r,r+1}=\,& & - {1 \over 2}\bigl[ J + \alpha (u_r -
u_{r+1})\bigr] \label{1}
\end{eqnarray}

where $J$ is the nearest neighbors hopping integral for an
undistorted chain, $\alpha$ is the electron-phonon coupling, $u_r$
is the dimerization coordinate which specifies the displacement of
the monomer group on the $r-$ lattice site along the molecular
axis, $f^{\dag}_r$ and $f_{r}$ create and destroy electrons (i.e.,
$\pi$ band electrons in polyacetylene) on the $r-$ group. Our
non-interacting Hamiltonian is given by a set of independent
oscillators which will be treated classically in the following.
The real space Hamiltonian in eq.(1) can be transformed in a time
dependent Hamiltonian by introducing $x(\tau)$ and $y(\tau')$ as
the electron coordinates at the $r$ and $r+1$ lattice sites,
respectively. The spatial {\it e-ph} correlations contained in
eq.(1) are then mapped onto the time axis by changing: $ u_r \to
u(\tau)$ and $u_{r+1} \to u(\tau')$. Accordingly, the time
dependent Hamiltonian is:

\begin{eqnarray}
H(\tau,\tau')=\,& &  J_{\tau, \tau'} \Bigl(f^{\dag}(x(\tau))
f(y(\tau')) + f^{\dag}(y(\tau')) f(x(\tau)) \Bigr)\, \nonumber \\
J_{\tau, \tau'}=\,& & - {1 \over 2}\bigl[J + \alpha(u(\tau) -
u(\tau'))\bigr] \label{2}
\end{eqnarray}

The twofold degenerate ground state of the SSH Hamiltonian
undergoes a Peierls instability which opens a gap at the Fermi
energy. In real space the soliton connects the two degenerate
phases with different senses of dimerizations and a localized
electronic state is associated with each soliton. Both electron
hopping between solitons and electron hopping to band states
(thermal excitation) are in principle possible within the model.
Mapping the Hamiltonian onto the time scale we maintain these
properties while thermally activated electron hops become time
dependent. $H(\tau,\tau')$ is therefore more general than the real
space SSH Hamiltonian since hopping processes are not constrained
to first neighbors sites along the chain. We emphasize that a
variable  range hopping introduces some local disorder in the
system.

Eq.(2) displays the semiclassical nature of the model in which
quantum mechanical degrees of freedom interact with the classical
variables $u(\tau)$. Setting $\tau'=\,0$, $u(0)\equiv y(0) \equiv
0$ and averaging the electron operators over the ground state we
obtain the time dependent semiclassical energy which is linear in
the displacements:

\begin{eqnarray}
<H(\tau)>=& &\,-{1 \over {2}} \bigl(J + \alpha u(\tau)
\bigr)\Bigl(G[x(\tau), \tau] + G[-x(\tau), -\tau] \Bigr)\,
 \nonumber \\
G[x(\tau), \tau]=& &\,{L \over {\beta \hbar}}\int_{-\pi/a}^{\pi/a}
{{dk}\over {2\pi}} exp[ik x(\tau)]\sum_n {{exp(-i\nu_n \tau)}
\over {i\nu_n - \epsilon_k /\hbar}} \, \nonumber \\ \label{3}
\end{eqnarray}

where, $L$ is the chain length, $a$ is the lattice constant,
$\beta$ is the inverse temperature, $\nu_n$ are the fermionic
Matsubara frequencies and $\epsilon_k=\, -J\cos(k)$ is the
electron dispersion relation. The chemical potential has been
pinned to the zero energy level. After summing over the
frequencies in eq.(3) we observe that the average energy per site
can be written as

\begin{eqnarray}
& &{{<H(\tau)>} \over N}=\, V\bigl(x(\tau)\bigr) +
u(\tau)j(\tau)\,
 \nonumber \\
& &V\bigl(x(\tau)\bigr)=\,-J {a \over \pi} \int_{0}^{\pi/a}{dk}
\cos[k x(\tau)]  \cosh(\epsilon_k \tau /\hbar) n_F(\epsilon_k) \,
\nonumber \\ & &j(\tau)= -\alpha {a \over \pi}
\int_{0}^{\pi/a}{dk} \cos[k x(\tau)]  \cosh(\epsilon_k \tau
/\hbar) n_F(\epsilon_k) \, \nonumber \\ \label{4}
\end{eqnarray}

with $N=\,L/a$ being the number of lattice sites and $n_F$ is the
Fermi function. $V\bigl(x(\tau)\bigr)$ is an effective term
accounting for the $\tau$ dependent electronic hopping while
$j(\tau)$ is the external source \cite{kleinert} current for the
oscillator field $u(\tau)$. Averaging the electrons over the
ground state we neglect the fermion-fermion correlations
\cite{hirsch} which lead to effective polaron-polaron interactions
in non perturbative analysis of the model \cite{acqua}. This
approximation however is not expected to affect substantially the
following calculations.

Taking a large number of oscillators ($u_i(\tau), i=1..{\bar N}$)
as the {\it bath} for the quantum mechanical particle whose
coordinate is $x(\tau)$, one can write the general path integral
at any temperature as:

\begin{eqnarray}
& &<x(\beta)|x(0)>=\,\prod_i \int Du_i(\tau) \int Dx(\tau) \cdot
\, \nonumber \\ & &exp\Biggl[-{1 \over \hbar} \int_0^{\hbar \beta}
d\tau \sum_i {{M_i} \over 2} \Bigl(\dot{u_i}^2(\tau) + \omega_i^2
u_i^2(\tau) \Bigr) \Biggr] \cdot \, \nonumber \\ & &exp\Biggl[-{1
\over \hbar} \int_0^{\hbar \beta} d\tau \biggl({m \over 2}
\dot{x}^2(\tau) + V\bigl(x(\tau)\bigr) -
\sum_i\gamma_iu_i(\tau)j(\tau)\biggr) \Biggr] \, \nonumber
\\ \label{5}
\end{eqnarray}

where, $m$ is the electron mass, $M_i$ is the mass of the $i-th$
ionic group, $\omega_i$ is the oscillator frequency, $\gamma_i$ is
the coupling constant between external source term $j(\tau)$ and
$i-th$ oscillator $u_i$. In our model, $M_i \equiv M$ and
$\gamma_i=\,1$ hence, integrating out the oscillators coordinates
over the paths $Du_i(\tau)$, imposing a closure condition on the
paths and replacing $\tau \to \tau/\hbar$, we get

\begin{eqnarray}
& &<x(\beta)|x(0)>=\,\prod_i Z_i \int Dx(\tau) \cdot \, \nonumber
\\ & &exp\Biggl[- \int_0^{\beta}d\tau \biggl({m \over 2}
\dot{x}^2(\tau) + V\bigl(x(\tau)\bigr) \biggr) - {1 \over \hbar}
A(x(\tau)) \Biggr], \, \nonumber \\ & &A(x(\tau))=\, -{{\hbar^2}
\over {4M}}\sum_{i=1}^{\bar N} {1 \over {\hbar \omega_i
\sinh(\hbar\omega_i\beta/2)}}\cdot \, \nonumber \\ & &
\int_0^{\beta} d\tau j(\tau) \int_0^{\beta}d{\tau''}
\cosh\Bigl(\omega_i \bigl( |\tau - {\tau''}| - \beta/2 \bigr)
\Bigr) j({\tau''}), \, \nonumber
\\ & &Z_i=\, {1 \over {2\sinh(\hbar\omega_i\beta/2)}} \label{6}
\end{eqnarray}

The quadratic source term $A(x(\tau))$ is the non local (in time)
action for the particle having memory of the interactions with the
oscillators field. Assuming periodic conditions $x(\tau)=\,x(\tau
+ \beta)$, the particle paths can be expanded in Fourier
components

\begin{eqnarray}
& &x(\tau)=\,x_o + \sum_{n=1}^\infty 2\Bigl(\Re x_n \cos( \omega_n
\tau) - \Im x_n \sin( \omega_n \tau) \Bigr)\, \nonumber \\ &
&\omega_n=\,2\pi n/\beta \label{7}
\end{eqnarray}

and, taking the following measure of integration

\begin{equation}
\oint Dx(\tau)\equiv \int_{-\infty}^{\infty}{{dx_o} \over
{\sqrt{2\pi\hbar^2/mK_BT}}} \prod_{n=1}^{\infty}
\Biggl[{{\int_{-\infty}^{\infty} d\Re x_n \int_{-\infty}^{\infty}
d\Im x_n} \over {\pi \hbar^2 K_BT/m\omega_n^2}} \Biggr]  \label{8}
\end{equation}

a direct integration of eq.(6) can be carried out in order to
derive the full partition function of the system versus
temperature.

Let's point out that, by mapping the electronic hopping motion
onto the time scale, a continuum version of the interacting
Hamiltonian (eq.(2)) has been de facto introduced. Unlike previous
\cite{lu} approaches however, our path integral method is not
constrained to the weak {\it e-ph} coupling regime and it can be
applied to any range of physical parameters.

\section*{II. Computational Method and Results}

As a preliminar step we determine, for a given path and at a given
temperature: i) the minimum number ($N_k$) of $k-$points in the
Brillouin zone to accurately estimate the average interacting
energy per lattice site and, ii) the minimum number ($N_{\tau}$)
of points in the double time integration to get a numerically
stable source action in eq.(6). We find: $N_k=\,70$ and
$N_{\tau}=\,300$ at $T=1K$.

Computation of eqs.(6)-(8) requires fixing two sets of input
parameters. The first set contains the physical quantities
characterizing the system: the bare hopping integral $J$, the
oscillator frequencies $\omega_i$ and the effective coupling
$\chi=\, \alpha^2 \hbar^2 /M$ (in units $meV^3$). The second set
defines the paths for the particle motion which mainly contribute
to the partition function through: the number of pairs ($\Re x_n,
\Im x_n$) in the Fourier expansion of eq.(7), the cutoff
($\Lambda$) on the integration range of the expansion coefficients
in eq.(8) and the related number of points ($N_\Lambda$) in the
measure of integration which ensures numerical convergence.

After introducing a dimensionless path

\begin{equation}
x(\tau)/a=\,{\bar x_o} + \sum_{n=1}^{N_p} \Bigl(\bar a_n \cos(
\omega_n \tau) + \bar b_n \sin( \omega_n \tau) \Bigr), \label{9}
\end{equation}

with: $\bar a_n \equiv 2 \Re x_n /a$ and $\bar b_n \equiv -2 \Im
x_n /a$, the functional measure of eq.(8) can be rewritten for
computational purposes as:

\begin{eqnarray}
\oint Dx& &(\tau) \approx {a \over {\sqrt{2}}}\biggl({a \over
2}\biggr)^{2N_p} {{(2\pi \cdot 4\pi \cdot \cdot \cdot 2N_p \pi)^2}
\over { (\pi\hbar^2/mK_BT)^{N_p + 1/2}}} \int_{-\Lambda
/a}^{\Lambda /a}{d \bar x_o} \cdot \, \nonumber \\ &
&\int_{-2\Lambda /a}^{2\Lambda /a}{d \bar a_1} \int_{-2\Lambda
/a}^{2\Lambda /a}{d \bar b_1} \cdot \cdot \cdot \int_{-2\Lambda
/a}^{2\Lambda /a}{d \bar a_{N_p}} \int_{-2\Lambda /a}^{2\Lambda
/a}{d \bar b_{N_p}}\, \nonumber \\ \label{10}
\end{eqnarray}

We take the lattice constant $a=\,1 \AA$. As a criterion to set
the cutoff $\Lambda$ on the integration range, we notice that the
functional measure normalizes the kinetic term in eq.(6):

\begin{equation}
\oint Dx(\tau) exp\Bigl[-{m \over 2}\int_0^\beta d\tau
\dot{x}^2(\tau)\Bigr]\equiv 1 , \label{11}
\end{equation}

and this condition holds for any number of pairs $N_p$ truncating
the Fourier expansion in eq.(9). Then, taking $N_p=\,1$ and using
eq.(10), the left hand side transforms as:

\begin{eqnarray}
& &\oint Dx(\tau) exp\Bigl[-{m \over 2}\int_0^\beta d\tau
\dot{x}^2(\tau)\Bigr] \simeq {4 \over \pi} \Bigl[\int_{0}^U dy
exp(-y^2)\Bigr]^2 \, \nonumber \\ & &U \equiv \sqrt{2
\pi^3}{\Lambda \over \lambda} \, \nonumber \\ & & \lambda=\,
\sqrt{{2 \pi \hbar^2} \over {m K_BT}}. \label{12}
\end{eqnarray}

Using the series representation \cite{grad}

\begin{equation}
\int_{0}^U dy exp(-y^2)=\,\sum_{k=0}^\infty {{(-1)^k U^{2k+1}}
\over {k! (2k+1)}},  \label{13}
\end{equation}

and taking $U=\,3$ (the series converges with $k_{max}\sim \,100$)
eq.(13) yields 0.886207 which well approximates the Poisson
integral value $\sqrt{\pi}/2$.

Thus, the cutoff $\Lambda$ can be expressed in terms of the
thermal wavelength $\lambda$ as $\Lambda \sim 3 \lambda / \sqrt{2
\pi^3}$ hence, it scales versus temperature as $\Lambda \propto
1/\sqrt{T}$. This means physically that, at low temperatures,
$\Lambda$ is large since many paths are required to yield the
correct normalization. For example, at $T=\,1K$, we get $\Lambda
\sim 284 \AA$. Numerical investigation of eq.(6) shows however
that a much shorter cutoff suffices to guarantee convergence in
the path integral, while the cutoff temperature dependence implied
by eq.(12) holds also in the computation of the interacting
partition function. The thermodynamical results hereafter
presented have been obtained by taking $\Lambda \sim \lambda /(10
\sqrt{2 \pi^3})$. Summing over $N_\Lambda \sim 20 /\sqrt{T}$
points for each integration range and taking $N_p=\,2$, we are
then evaluating the contribution of $(N_\Lambda + 1)^{2N_p + 1}$
paths (the integer part of $N_\Lambda$ is obviously selected at
any temperature). Thus, at $T=\,1K$ we are considering $\sim 4
\cdot 10^6$ different paths for the particle motion while, at
$T=\,100K$ the number of paths drops to 243. Low temperature
calculations prove therefore to be extremely  time consuming. Note
that larger $N_p$ in the path Fourier expansion would further
increase the computing time without introducing any substantial
improvement in the thermodynamical output of our calculation.

Although the history of the SSH model is mainly related to wide
band polymers, we take here a narrow band system ($J=\,100meV$) to
be consistent with previous investigations \cite{io} and with the
caveat that electron-electron correlations may become relevant in
narrow bands. Free energy and heat capacity have been first
(Figures 1-3) computed up to room temperature assuming a bath of
$\bar N=\,10$ low energy oscillators separated by $2meV$: $\hbar
\omega_1=\,2meV,..., \hbar \omega_{10}=\,20meV$. The lowest energy
oscillator yields the largest contribution to the phonon partition
function mainly in the low temperature regime while the
$\omega_{10}$ oscillator essentially sets the phonon energy scale
which determines the size of the {\it e-ph} coupling. A larger
number $\bar N$ of oscillators in the aforegiven range would not
significantly modify the calculation.

In the discrete SSH model, the value $\bar \alpha \equiv \,
4\alpha^2/(\pi \kappa J) \sim 1$, marks the crossover between weak
and strong {\it e-ph} coupling, with $\kappa$ being the effective
spring constant. In our continuum and semiclassical model the
effective coupling is the above defined $\chi$. Although in
principle, discrete and continuum models may feature non
coincident crossover parameters, we assume that the relation
between $\alpha$ and $J$ obtained by the discrete model crossover
condition still holds in our model. Hence, at the crossover we
get: $\chi_c \sim \, \pi J \hbar^2 \omega^2_{10}/64$. This means
that, in Figures 1-3, the crossover value is set at $\chi_c \sim
\,2000meV^3$.

In Fig.1, the phonon free energy ($F_{ph}$) is plotted versus
temperature together with the free energy due to the total action
in eq.(6). $F_{sou}$ results from the competition between the free
path action (kinetic term plus hopping potential in the
exponential integrand) and the source action depending on the {\it
e-ph} coupling. While the former enhances the free energy the
latter becomes dominant at increasing temperatures thus reducing
the total free energy. As a main feature one notes that, by {\it
increasing} $\chi$, $F_{sou}$ gets a negative derivative at {\it
decreasing} temperatures. In the very weak coupling case
($\chi=\,100meV^3$) $F_{sou}$ never intersects $F_{ph}$ whereas an
intersection point shows up both at moderately weak
($\chi=\,1000meV^3$) and at moderately strong ($\chi=\,3000meV^3$)
couplings. The intersection temperature decreases as expected by
increasing the strength of $\chi$ but, at low $T$ ($\prec 50K$),
the contribution of the free electron path action prevails: this
feature is reflected on the low $T$ behavior of the heat capacity
linear coefficient hereafter discussed.

Fig.2 shows the source term contributions to the heat capacity:
the previously described summation over a large number of paths
turns out to be essential to recover the correct thermodynamical
behavior in the zero temperature limit. The small phonon
contribution to the heat capacity is plotted in Fig.3 to point out
that the Dulong Petit value is achieved at $T \succeq 200K$.
Looking at the {\it total heat capacity over temperature} ratio,
we find a peculiar low temperature upturn (also in the weak $\chi$
regime) which can be mainly ascribed to the sizable effective
hopping integral term $V\bigl(x(\tau)\bigr)$. The {\it e-ph}
coupling however determines the shape of the low $T$ anomaly.

The oscillators bath affects the electronic correlations both on
the space and time scales. Analysis of the thermal correlation
function shows infact that high energy phonons substantially
reduce the electronic correlation length at low temperatures
while, in presence of low energy phonons, the electronic paths are
correlated over the $\tau$ scale. By increasing $T$, the role of
the phonons on the correlation function is less pronounced.

The effect of the oscillators bath on the thermodynamical
properties is discussed in Figures 4-6 where the ten phonon
energies are: $\hbar \omega_1=\,22meV,..., \hbar
\omega_{10}=\,40meV$. Accordingly the crossover is set at $\chi_c
\sim \,8000meV^3$ and three plots out of five lie in the strong
{\it e-ph} coupling regime. As shown in Fig.4, much larger $\chi$
values (with respect to Fig.1) are required to get strongly
decreasing free energies versus temperature while the
$\chi=\,3000meV^3$ curve now hardly intersects the phonon free
energy at room temperature. Fig.5 shows the rapid growth of the
source heat capacity versus temperature at strong couplings
whereas the presence of the low $T$ upturn in the {\it total heat
capacity over T} ratio is confirmed in Fig.6. Note that, due to
the enhanced oscillators energies, the phonon heat capacity
saturates at $T \sim 400K$.

According to our integration method (eq.(10), at any temperature,
a specific set of Fourier coefficients defines the ensamble of
relevant particle paths over which the hopping potential
$V\bigl(x(\tau)\bigr)$ is evaluated. This ensamble is therefore
$T$ dependent. However, given a single set of path parameters one
can monitor the $V\bigl(x(\tau)\bigr)$ behavior versus $T$. It
turns out that the hopping decreases (as expected) by lowering $T$
but its value remains appreciable also at low temperatures
($\preceq 20K$). Since the $d\tau$ integration range is larger at
lower temperatures, the overall hopping potential contribution to
the total action is relevant also at low $T$. It is precisely this
property which is responsible for the anomalous  upturn in the
heat capacity linear coefficient. Further investigation also
reveals that the upturn persists both in the extremely narrow ($J
\sim 10meV$) and in the wide band ($J \sim 1eV$) regimes. Our
computation method accounts for a variable range hopping on the
$\tau$ scale which corresponds physically to introduce some degree
of disorder along the linear chain. This feature makes our model
more general than the standard SSH Hamiltonian (eq.(1)) with only
real space nearest neighbor hops. While hopping type mechanisms
have been suggested \cite{kivel} to explain the striking
conducting properties of doped polyacetylene at low temperatures
we are not aware of any direct computation of specific heat in the
SSH model. Since the latter quantity directly probes the density
of states and integrating over $T$ the {\it specific heat over T}
ratio one can have access to the experimental entropy, our method
may provide a new approach to analyse the transition to a
disordered state which indeed exists in polymers \cite{nieu}. In
this connection it is also worth noting that the low $T$ upturn in
the specific heat over $T$ ratio is a peculiar property of glasses
\cite{zeller,anderson} in which tunneling states for atoms (or
group of atoms) provide a non magnetic internal degree of freedom
in the potential structure \cite{io91}.

\begin{figure}
\vspace*{8truecm} \caption{Phonon and Source Term contributions to
the free energy for three values of the effective coupling $\chi$
(in units $meV^3$) and a narrow electron band. A bath of ten
phonon oscillators is considered, the largest phonon energy being
$\hbar \omega_{10}=\,20meV$. }
\end{figure}

\begin{figure}
\vspace*{8truecm} \caption{Source Term contributions to the heat
capacity for the same parameters as in Fig.1. }
\end{figure}

\begin{figure}
\vspace*{8truecm} \caption{Total heat capacity over temperature
for the same parameters as in Fig.1. The phonon heat capacity is
also plotted.}
\end{figure}

\begin{figure}
\vspace*{8truecm} \caption{Phonon and Source Term contributions to
the free energy for five values of the effective coupling $\chi$
(in units $meV^3$) and a narrow electron band. A bath of ten
phonon oscillators has been taken, the largest phonon energy is
$\hbar \omega_{10}=\,40meV$. }
\end{figure}

\begin{figure}
\vspace*{8truecm} \caption{Source Term contributions to the heat
capacity for the same parameters as in Fig.4. }
\end{figure}

\begin{figure}
\vspace*{8truecm} \caption{Total heat capacity over temperature
for the same parameters as in Fig.4. The phonon heat capacity is
also plotted.}
\end{figure}

\section*{III. Conclusions}

Mapping the real space Su-Schrieffer-Heeger model onto the time
space we have developed a semiclassical version of the interacting
model Hamiltonian which can be attacked by path integrals methods.
The acoustical phonons of the standard SSH model have been
replaced by a set of oscillators providing a bath for the electron
interacting with the displacements field. Time retarded
interactions are naturally introduced in the formalism through the
source action $A(x(\tau))$ which depends quadratically on the bare
{\it e-ph} coupling strength $\alpha$. Via calculation of the
electronic motion path integral, the partition function can be
derived in principle for any value of $\alpha$ thus avoiding those
limitations on the {\it e-ph} coupling range which burden any
perturbative method. Particular attention has been paid to
establish a reliable and general procedure which allows one to
determine those input parameters intrinsic to the path integral
formalism. It turns out that a large number of paths is required
to carry out low temperature calculations which therefore become
highly time consuming. The physical parameters have been specified
to a narrow band system and the behavior of some thermodynamical
properties, free energy and heat capacity, has been analysed for
some values of the effective coupling strength lying both in the
weak and in the strong coupling regime. We find a peculiar upturn
in the low temperature plots of the heat capacity over temperature
ratio indicating that a glassy like behavior can arise in the
linear chain as a consequence of a time dependent electronic
hopping with variable range.

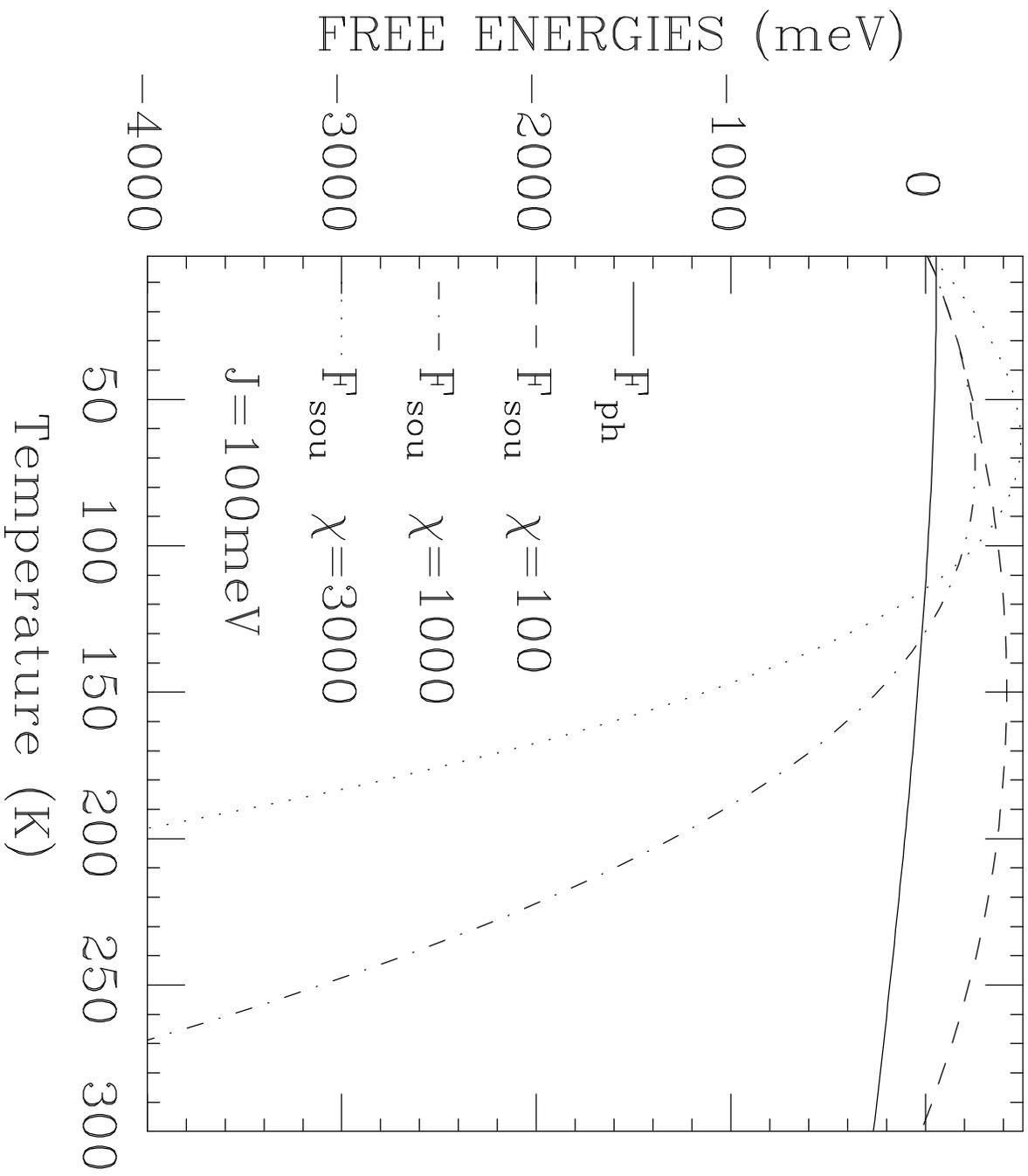
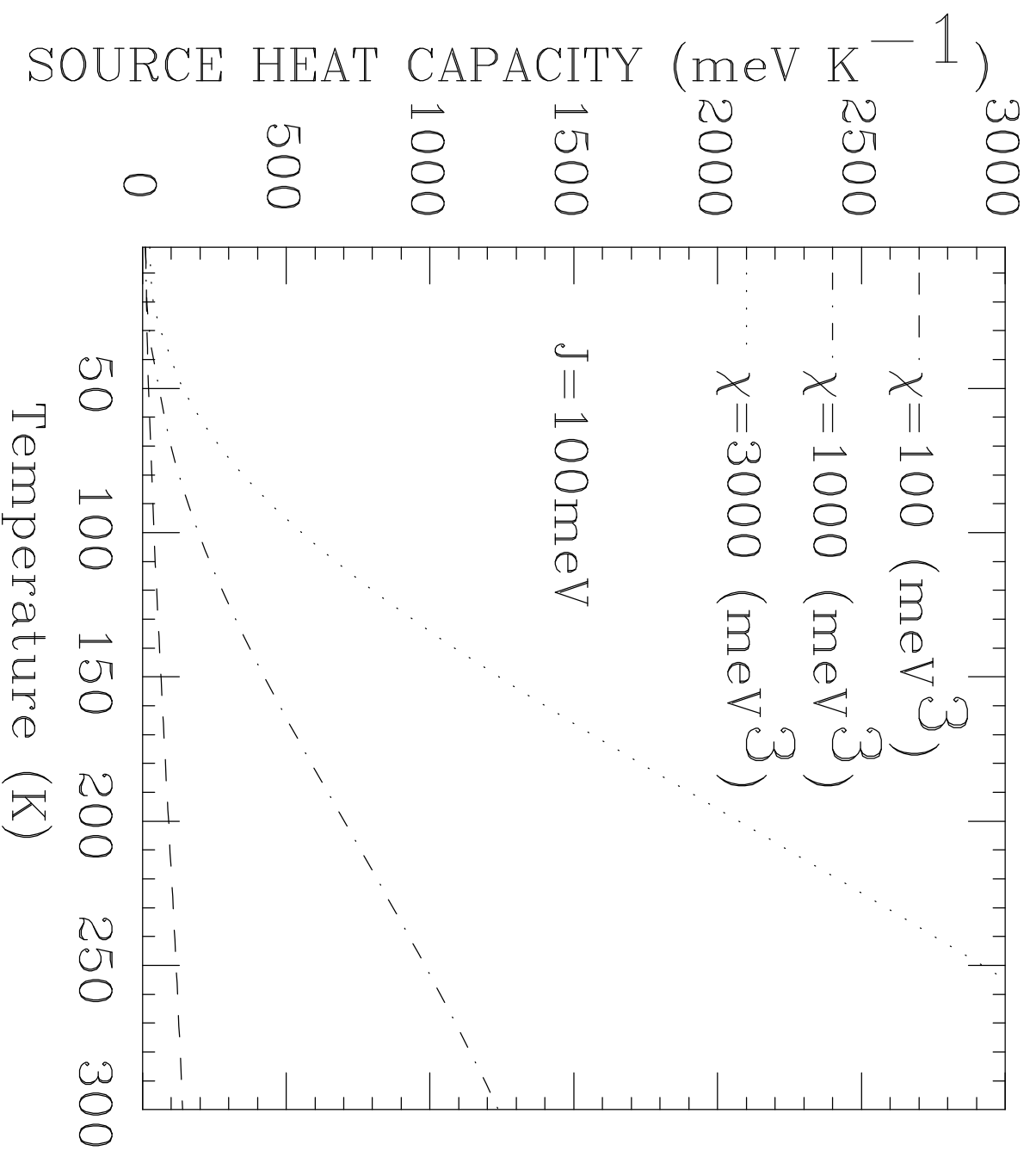
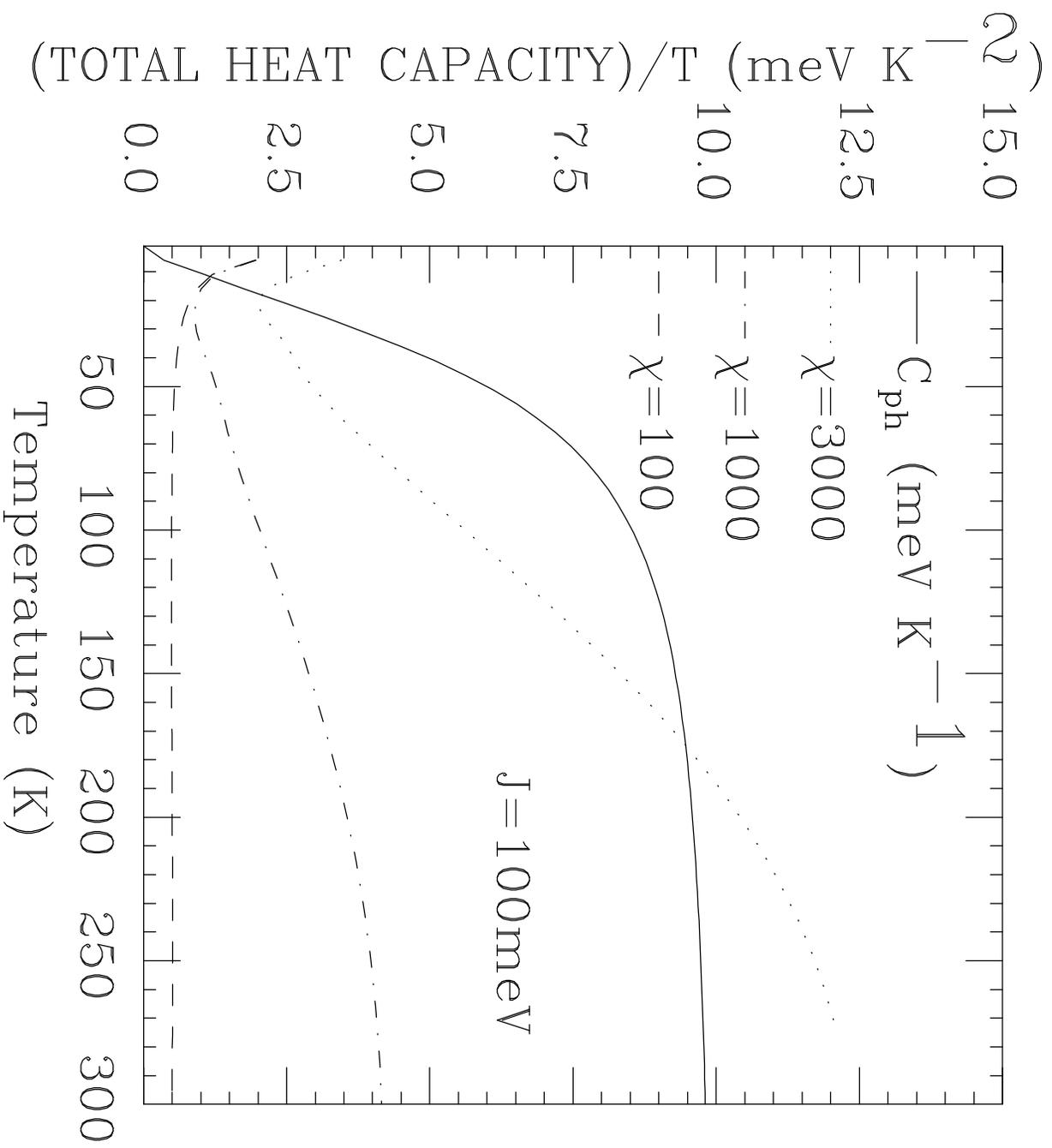
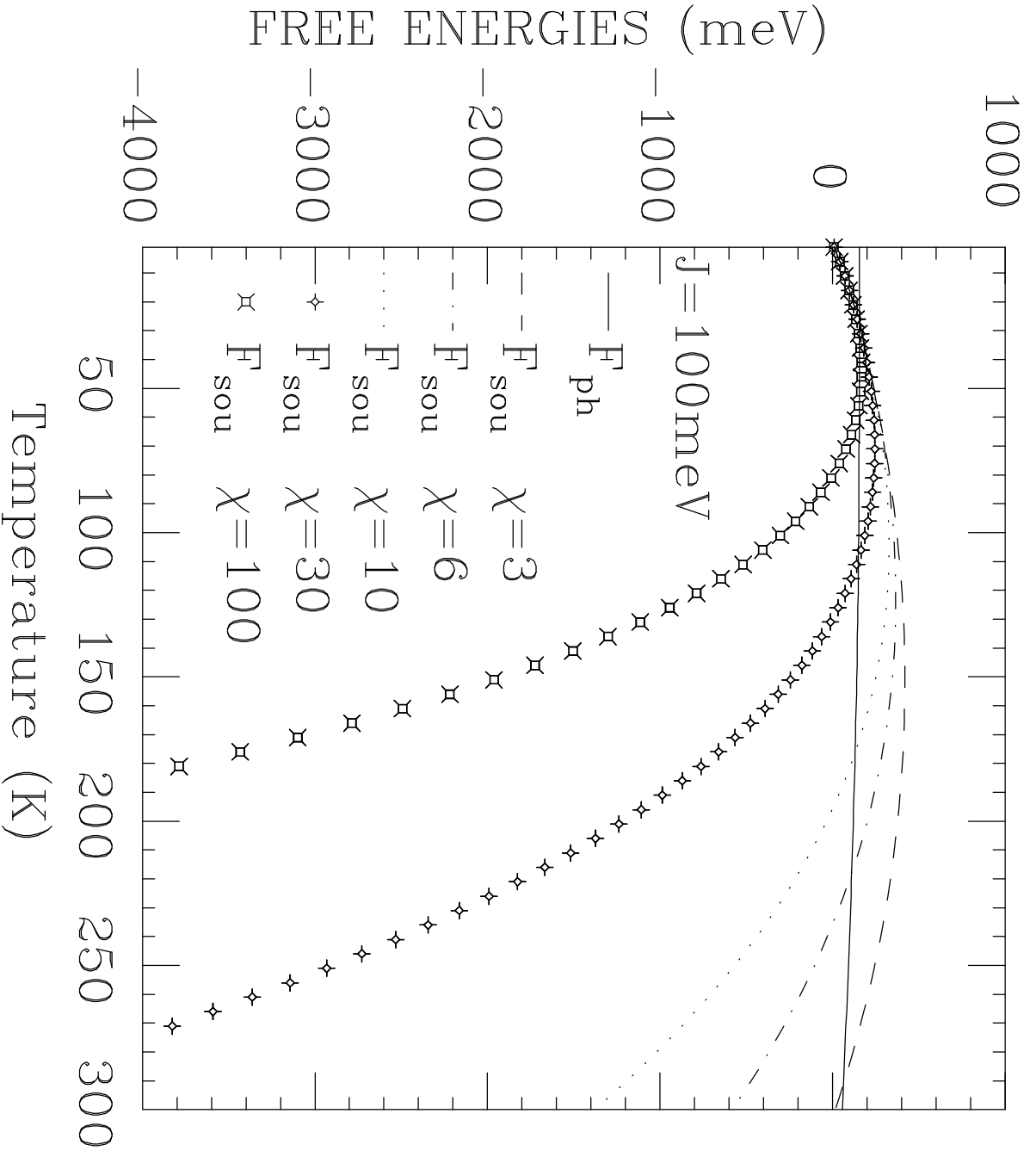
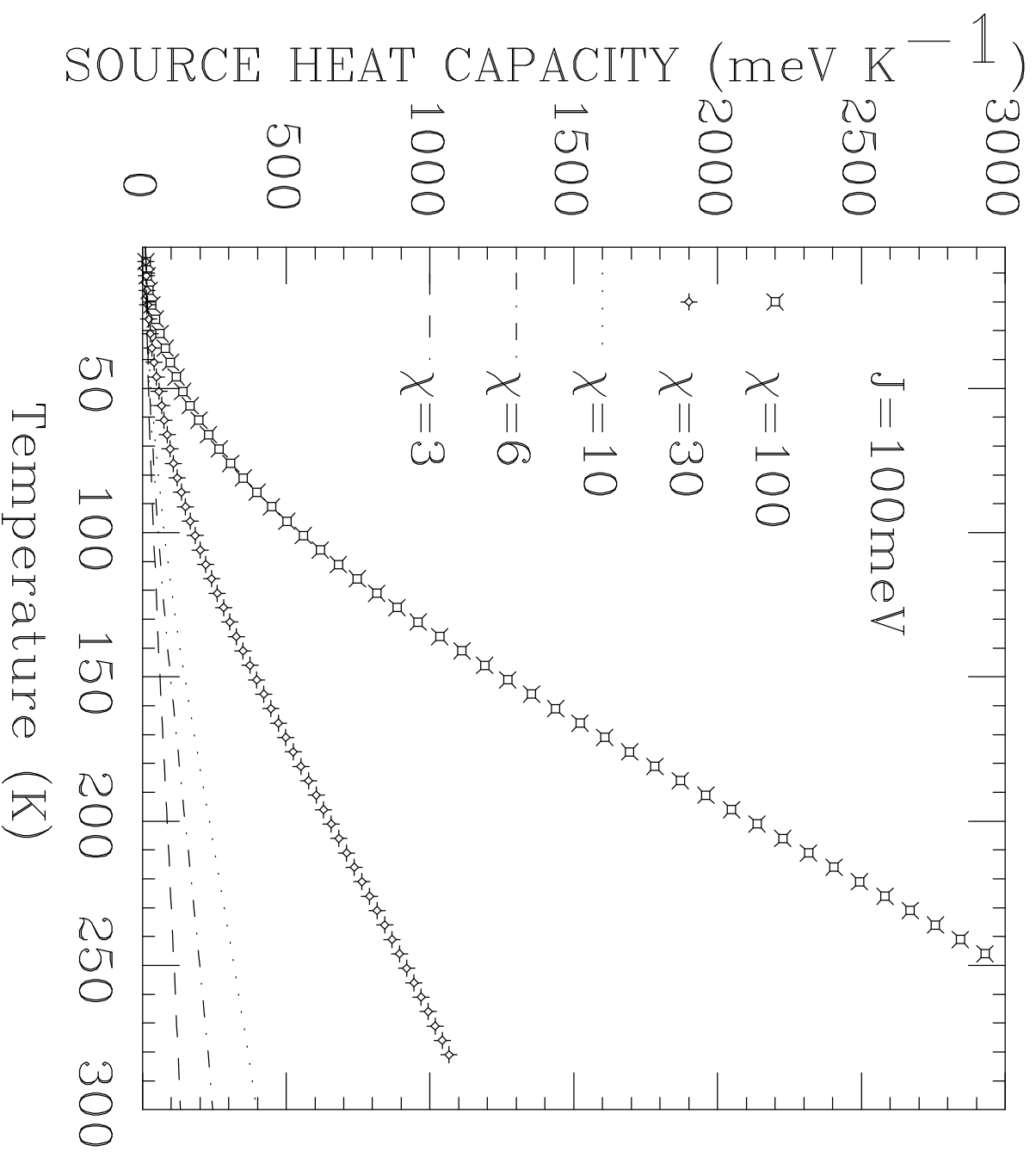
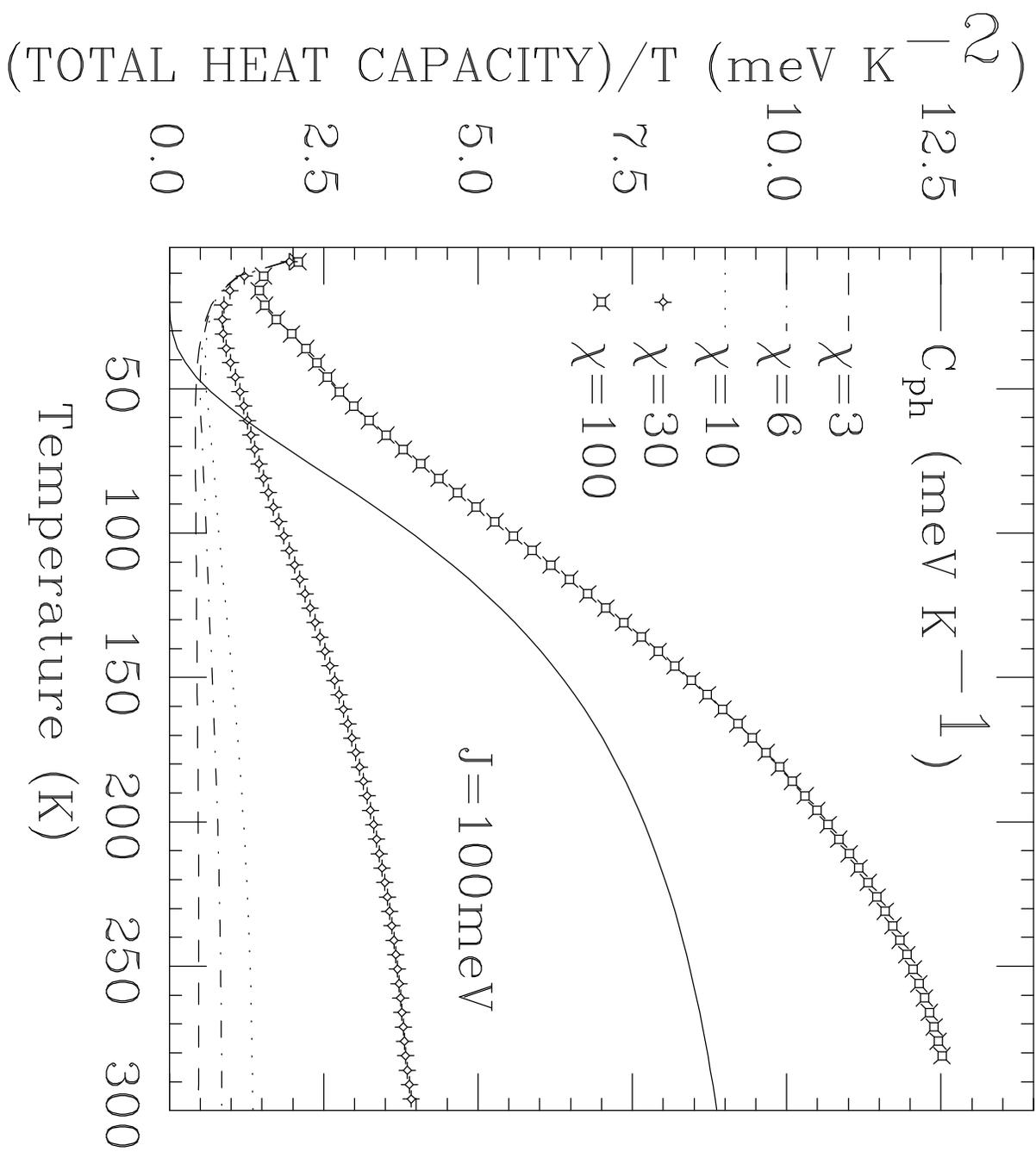
\end{document}